\documentclass[12pt]{article}
\usepackage{amsmath}
\usepackage{epsfig}
\usepackage{amssymb}
\usepackage{hyperref}
\usepackage{bbold}
\usepackage{braket}
\input{epsf}
\def\be{\begin{equation}}
\def\ee{\end{equation}}
\def\beq{\begin{equation}}
\def\eeq{\end{equation}}
\def\beqa{\begin{eqnarray}}
\def\eeqa{\end{eqnarray}}
\def\ba{\begin{eqnarray}}
\def\ea{\end{eqnarray}}
\def\bea{\begin{eqnarray}}
\def\eea{\end{eqnarray}}

\newcommand\J{{\mathrm{J}}}
\def\nn{\nonumber}

\begin{document}

\begin{titlepage}
\renewcommand{\thefootnote}{\fnsymbol{footnote}}
\par \vspace{10mm}

\begin{center}
\vspace*{10mm}
{\large \bf
Next-to-leading Order Calculation for Jets Defined \\[4mm] 
by a Maximized Jet Function}
\end{center}

\par \vspace{2mm}
\begin{center}
{\bf Tom Kaufmann${}^{\,a}$,}
\hskip .2cm
{\bf Asmita Mukherjee${}^{\,b}$,}
\hskip .2cm
{\bf Werner Vogelsang${}^{\,a}$  }\\[5mm]
\vspace{5mm}
${}^{a}$ Institute for Theoretical Physics, T\"ubingen University, 
Auf der Morgenstelle 14, \\ 72076 T\"ubingen, Germany\\[2mm]
${}^{b}$ Department of Physics, Indian Institute of Technology Bombay, \\
Powai, Mumbai 400076, India
\end{center}


\vspace{9mm}
\begin{center} {\large \bf Abstract} \end{center}
We present a next-to-leading order QCD calculation for the single-inclusive production of collimated jets 
at hadron colliders, when the jet is defined by maximizing a suitable jet function that depends on the 
momenta of final-state particles in the event. A jet algorithm of this type was initially proposed by Georgi 
and subsequently further developed into the class of ``$J_{E_T}$ algorithms''. Our calculation establishes 
the infrared safety of the algorithms at this perturbative order. We derive analytical results for the relevant 
partonic cross sections. We discuss similarities and differences with respect to jets defined by cone or 
(anti-)$k_t$ algorithms and present numerical results for the Tevatron and the LHC. 

\end{titlepage}  

\setcounter{footnote}{2}
\renewcommand{\thefootnote}{\fnsymbol{footnote}}


\section{Introduction \label{intro}}

Jets produced at high-energy hadron colliders play important roles as precision probes of 
QCD and nucleon structure. They are also involved in many signal or background processes 
for searches for physics beyond the standard model. As a result, much work has gone into
the development of useful definitions of a jet~\cite{soyez}. An important prerequisite is that the 
jet definition is both viable for experimental studies and infrared safe in the theoretical calculation. 
The definitions used so far can be broadly divided into two classes: cone algorithms and successive 
recombination algorithms. The ``Seedless Infrared Safe Cone'' (SISCone) algorithm~\cite{soyez1} 
on the one hand, and the $k_t$~\cite{ES,catani} and anti-$k_t$~\cite{css} algorithms on the other hand
are important representatives of these two classes of jet algorithms that fulfill the desired criteria. 

Recently, a further method for defining jets was proposed by Georgi~\cite{georgi}. 
Here the idea is to assemble jets by maximizing a suitable function of the four-momenta of final-state
particles. Defining the total four-momentum of a given subset of the final-state particles as
\be
P_{\mathrm{\,set}}\,\equiv\,\sum_{i\,\in\,\mathrm{set}} p_i\,,
\ee
this function may in its simplest form be defined as
\be\label{Jdef}
J(P_{\mathrm{\,set}})\,\equiv\,E_{\mathrm{\,set}}^\perp - \beta \,
\frac{m_{\mathrm{\,set}}^2}{E_{\mathrm{\,set}}^\perp}\;,
\ee
where $m_{\mathrm{\,set}}^2\equiv (P_{\mathrm{\,set}})^2$.
This definition corresponds to the one originally given in~\cite{georgi}, except that we have 
followed Ref.~\cite{Bai:2014qca} to use the total {\it transverse} energy, $E_{\mathrm{\,set}}^\perp \equiv
\sqrt{(P_{\mathrm{\,set}}^{\,x})^2+(P_{\mathrm{\,set}}^{\,y})^2+m_{\mathrm{\,set}}^2}$, of the particles 
rather than just their energy, the former being boost-invariant and hence more appropriate for the application to 
hadronic scattering. In this version, the algorithm has been termed ``$J_{E_T}$ algorithm'' in~\cite{Bai:2014qca}.
The parameter $\beta>1$ is fixed and specifies the algorithm. The idea behind maximizing $J$ 
in Eq.~(\ref{Jdef}) is that it forces particles to be arranged in collimated jets: if the invariant mass 
$m_{\mathrm{\,set}}$ of the set is large, the set will not produce a global maximum of $J$. So only
high-$E_{\mathrm{\,set}}^\perp$, low-mass, sets give rise to jets. A reconstructed jet thus 
maximizes the function $J$ with the value
\be
J(P_\J)\,\equiv\,E^\perp_\J - \beta \frac{m_\J^2}{E^\perp_\J}\;,
\ee
$P_\J$ and $E^\perp_\J$ being the four-momentum and transverse energy
of the jet, respectively, and $m_\J$ its invariant mass. The algorithm is iterative; once a jet has been found, 
its particles are removed from the list of particles in the event, and the algorithm is applied to the remaining ones.  
We note that variants of the function $J$ may be introduced, which may potentially improve the applicability
in actual experimental jet analyses. For instance, one can define a class of $J_{E_T}$ algorithms by considering
weighted functions~\cite{Bai:2014qca,Ge:2014ova}
\be\label{Jdefn}
J^{(n)}(P_{\mathrm{\,set}})\,\equiv\,(E_{\mathrm{\,set}}^\perp)^n 
\left(1- \beta \frac{m_{\mathrm{\,set}}^2}{(E_{\mathrm{\,set}}^\perp)^2}\right)
\ee
for the maximization procedure, where the case $n=1$ corresponds to~(\ref{Jdef}), i.e. 
$J^{(1)}=J$. Other choices are conceivable\footnote{For instance, one could define
$\tilde{J}^{(k)}(P_{\mathrm{\,set}})\,\equiv\,E_{\mathrm{\,set}}^\perp
(1- \beta \,(m_{\mathrm{\,set}}/E_{\mathrm{\,set}}^\perp)^{2k})$, $k>0$,
which increases or decreases the ``penalty'' for sets with large $m_{\mathrm{\,set}}$.
Although infrared-safe at NLO, this choice modifies the infrared structure of the cross section by
changing the threshold logarithms, which is not desirable.}. We also note that related ideas for
defining jets were introduced earlier in the context of the ``$N$-jettiness'' 
observable~\cite{Stewart:2010tn}.

In this paper, we will present next-to-leading order (NLO) results for single-inclusive
jet production for some of these new jet algorithms. Specifically, we will address the 
cases $n=1$ and $n=2$ in~(\ref{Jdefn}). Following the earlier work~\cite{Aversa:1990ww,jet1,jet2}, 
we will derive analytical results for the partonic NLO cross sections by assuming that 
the jets are rather collimated. This is an approximation that was termed ``Narrow Jet 
Approximation'' (NJA) in~\cite{jet2}. In the context of the novel jet algorithms defined
by Eqs.~(\ref{Jdef}),(\ref{Jdefn}) it means that $\beta$ is chosen to be
rather large, $\beta \gg 1$. In fact, as we shall see, $\beta$ very closely corresponds to 
$1/R^2$ where $R$ is the jet parameter in the more standard jet definitions, i.e. 
the cone opening in cone algorithms or the ``distance'' parameter in (anti-)$k_t$ algorithms.
In perturbation theory, the NLO jet cross section for the new algorithms will exhibit a form 
${\cal A}\log (1/\beta) +{\cal B}+{\cal O} (1/\beta)$. The coefficients ${\cal A}$ and 
${\cal B}$ are determined analytically in our NJA approach. As was shown in~\cite{jet1,jet2}, 
for the standard jet algorithms the NJA is very accurate even at relatively large $R\sim 0.7$. 
For our present study, this implies that our calculations will be accurate even for values of $\beta$ 
rather close to unity. 

The remainder of this paper is organized as follows: In Sec.~\ref{tech} we
present the technical details and analytical results of our calculation of single-inclusive
jet cross sections in the NJA for the new algorithms. Section~\ref{Pheno} contains a few
simple phenomenological results where we also compare to results for the more standard
algorithms. Section~\ref{con} contains our conclusions.

\section{Jet production at next-to-leading order in the NJA \label{tech}}

We consider single-inclusive jet production in hadronic collisions, for example $pp\to {\mathrm{jet}}X$.
To carry out the analytical NLO calculation, we follow~\cite{Aversa:1990ww,jet1,jet2} and apply proper modifications 
to the NLO cross sections $d\sigma_c$ for single-inclusive production of a specific {\it parton} $c=q,\bar{q},g$,
which were derived in~\cite{oldsca2,jssv}. This cross section is not by itself the desired jet cross section, 
because it has been integrated over the full phase space of all final-state partons other than $c$. Therefore, it
contains contributions where a second parton in the final state is so close to parton $c$ that the two should
jointly form the jet for a given jet definition. One can correct for this by subtracting 
such contributions from $d\sigma_c$ and adding a piece where they actually do form the jet together. At 
NLO, where there can be three partons $c,d,e$ in the final state, one has after suitable summation over all
possible configurations:
\begin{eqnarray}\label{deco}
d \sigma_{ab\rightarrow \mathrm{jet}X} &=&
[d \sigma_c -d \sigma_{c(d)}-
d \sigma_{c(e)}]+ [d \sigma_d -d \sigma_{d(c)}-
d \sigma_{d(e)}]+
[d  \sigma_e -d \sigma_{e(c)}-
d \sigma_{e(d)}]\nonumber\\[2mm] &+&
d \sigma_{cd} + d \sigma_{ce}+
d \sigma_{de} \, .
\label{jetform}
\end{eqnarray}
Here $d \sigma_j$ is again the single-parton inclusive cross section where parton $j$
is observed (which also includes the virtual corrections),  $d \sigma_{j(k)}$ is the cross section
where parton $j$ produces the jet, but parton $k$ is so close that it should be part of the jet,
and $d \sigma_{jk}$ is the cross section when both partons $j$ and $k$ jointly form the jet.
Final-state collinear singularities in $d \sigma_{j(k)}$ must be subtracted in the same way as they
were subtracted in the original calculation of the $d \sigma_j$. This guarantees that the two 
subtractions effectively cancel in the full jet cross section, as they must, since for a well-defined 
jet no subtraction should be necessary at all. 

While the decomposition~(\ref{deco}) is completely general to NLO, the $d \sigma_{j(k)}$ and 
$d \sigma_{jk}$ may be computed even analytically within the NJA. At NLO, they 
both receive contributions from real-emission $2\to 3$ diagrams only. For the NJA one assumes
that the observed jets are rather collimated. The relevant calculations for the standard cone and 
$k_t$ algorithms were carried out in Refs.~\cite{jet1,jet2}. It was observed that the $d \sigma_{j(k)}$ 
are the same for both types of algorithms, while the $d \sigma_{jk}$ differ by finite pieces. 
We will now apply the NJA to the jet definition obtained by maximizing the function $J^{(n)}$
in Eq.~(\ref{Jdefn}). 

We start with the computation of the $d\sigma_{jk}$. We consider the cross section differential in 
the variables 
\ba
v= 1+\frac{t}{s}\;,\;\; w= \frac{-u}{s+t}\;,
\label{partvw}
\ea
where the partonic Mandelstam variables are defined by
\be
s\equiv (p_a+p_b)^2,\;\;t\equiv (p_a-P_\J)^2,\;\;u\equiv(p_b-P_\J)^2,
\ee
with the momenta $p_a,p_b$ of the incoming partons. As was shown in~\cite{jet1,jet2}, in the NJA 
$d \sigma_{jk}$ is given by 
\beq
\label{eq:scaps5}
\frac{d\sigma_{jk}}{dvdw}\,=\,\frac{\alpha_s}{\pi}\,{\cal N}_{ab\to K}(v,w,\varepsilon)\,
\delta(1-w) \int_0^1 dz\, z^{-\varepsilon} (1-z)^{-\varepsilon}P_{jK}^{<}(z)
\,\int_0^{m^2_{{\mathrm{J,max}}}} \frac{dm_\J^2}{m_\J^2} \,m_\J^{-2 \varepsilon},
\eeq
where we have used dimensional regularization with $d=4-2\varepsilon$ space-time dimensions.
Equation~(\ref{eq:scaps5}) is derived from the fact that the leading contributions in the NJA come
from a parton $K$ splitting into partons $j$ and $k$ ``almost'' collinearly in the final state. We therefore have 
an underlying Born process $ab\to KX$ (with some unobserved recoil final state $X$), whose $d$-dimensional 
cross section is contained in the ``normalization factor'' ${\cal N}_{ab\to K}$, along with some trivial factors. 
The integrand then contains the $d$-dimensional splitting functions $P_{jK}^<(z)$,
where the superscript ``$<$'' indicates that the splitting function is strictly at $z<1$, that is, without
its $\delta(1-z)$ contribution that is present when $j=K$ (see~\cite{jet1,jet2}). The $P_{jK}^<(z)$ are
given by 
\ba\label{splits}
P_{qq}^<(z)&=&C_F \left[ \frac{1+z^2}{1-z}-\varepsilon (1-z)\right],\nn\\[2mm]
P_{qg}^<(z)&=&\frac{1}{2}\left[ z^2 + (1-z)^2 - 2 \varepsilon z (1-z)\right],\nn\\[2mm]
P_{gq}^<(z)&=&C_F \left[ \frac{1+(1-z)^2}{z}-\varepsilon z\right],\nn\\[2mm]
P_{gg}^<(z)&=&2 C_A \frac{(1-z+z^2)^2}{z (1-z)},
\ea
where $C_A=3$ and $C_F=4/3$. The argument $z$ of the splitting function is the fraction of the intermediate 
particle's momentum (equal to the jet momentum) transferred in the splitting. The second integral in~(\ref{eq:scaps5}) 
runs over the pair mass of partons $j$ and $k$, which for the contribution $d\sigma_{jk}$ is identical to the jet 
mass $m_\J$. The explicit factor $m_\J^2$ in the denominator represents the propagator of the splitting parton $K$. 
The integral over the pair mass of partons $j$ and $k$ runs between zero and an upper limit $m_{{\mathrm{J,max}}}$. 
In the NJA, where the two partons are assumed to be almost collinear, $m_{{\mathrm{J,max}}}$ is formally
taken to be relatively small, which justifies the approximations made in deriving Eq.~(\ref{eq:scaps5}). Note 
that powers of $m_\J^2$ have been neglected wherever possible, which makes the integral over $m_\J^2$ trivial.

The value of $m_{{\mathrm{J,max}}}^2$ depends on the jet algorithm chosen. We can straightforwardly 
derive it for the class of $J_{E_T}$ algorithms introduced by Eq.~(\ref{Jdefn}). For $\sigma_{jk}$, we just 
need to make sure that the two partons $j$ and $k$ really jointly form the jet. This requires that the value of 
the $J^{(n)}$ function constructed from the two partons together is larger than the value of $J^{(n)}$ for each 
parton individually, or
\beq
(E_\J^\perp)^n \left(1- \beta \frac{m_\J^2}{(E_\J^\perp)^2}\right)\geq \max\left((E_j^\perp)^n,(E_k^\perp)^n\right)\;,
\eeq
which implies
\beq\label{m11}
m_\J^2\leq  \frac{(E_\J^\perp)^2}{\beta} \left(1-\frac{\max\left((E_j^\perp)^n,(E_k^\perp)^n\right)}{(E_\J^\perp)^n}\right).
\eeq
From this it is evident that the NJA corresponds to the limit of large $\beta$. We note that in the NJA we may replace 
the transverse energies by the transverse momenta (denoted by $P_\J^\perp$ and $p_j^\perp$, $p_k^\perp$ for the 
jet and the partons, respectively),  since corrections introduced by this will always be suppressed
by an additional power of $1/\beta$. Using the relations $p_j^\perp=z P_\J^\perp$, $p_k^\perp=(1-z)P_\J^\perp$ 
appropriate for the splitting described above, Eq.~(\ref{m11}) finally gives
\beq
m_\J^2\leq\frac{(P_\J^\perp)^2}{\beta}\min\left(1-(1-z)^n,1-z^n\right)\;.
\eeq
The right-hand-side is the $m_{{\mathrm{J,max}}}^2$ we need. Inserting into~(\ref{eq:scaps5}) we arrive at 
\begin{eqnarray}
\label{eq:scaps5a}
\frac{d\sigma_{jk}}{dvdw}&=&\frac{\alpha_s}{\pi}\,{\cal N}_{ab\to K}(v,w,\varepsilon) \,
\delta(1-w)\,\left(-\frac{1}{\varepsilon}\right)\,\left( \frac{(P_\J^\perp)^2}{\beta} \right)^{-\varepsilon}
I_{jK}^{(n)}\;,
\end{eqnarray}
where
\beq
I_{jK}^{(n)}\,\equiv\,\left[\int_0^{1/2} dz  \, 
\left(1-(1-z)^n\right)^{-\varepsilon}+\int_{1/2}^1 dz \,
\left(1-z^n\right)^{-\varepsilon}\right]z^{-\varepsilon} (1-z)^{-\varepsilon}\,P_{jK}^{<}(z)\,.
\eeq
Evaluation of these integrals for general $n$ is tedious. In any case, we are only interested in the cases $n=1$ and $n=2$ 
here, for which the integrals may be easily computed in closed form. For $n=1$, expanding to ${\cal O}(\varepsilon)$ because
of the overall $1/\varepsilon$-pole in~(\ref{eq:scaps5a}), 
we find:
\bea\label{In1}
I_{qq}^{(1)}&=&C_F\left[-\frac{1}{\varepsilon}-\frac{3}{2} +\varepsilon \left(-5+\frac{\pi^2}{2}-\frac{3}{2}
\log 2\right)\right]\,=\,I_{gq}^{(1)}\;,\nn\\[2mm]
I_{qg}^{(1)}&=&\frac{1}{2}\left[\frac{2}{3}+\varepsilon\left(\frac{23}{12}+\frac{2}{3}\log 2\right)\right]\;,\nn\\[2mm]
I_{gg}^{(1)}&=&2 C_A\left[-\frac{1}{\varepsilon}-\frac{11}{6} +\varepsilon \left(-\frac{45}{8}+\frac{\pi^2}{2}-\frac{11}{6}
\log 2\right)\right]\;,
\eea
while for $n=2$
\bea\label{In2}
I_{qq}^{(2)}&=&C_F\left[-\frac{1}{\varepsilon}-\frac{3}{2} +\log 2+\frac{\varepsilon}{2} 
\left(-13+\pi^2+18 \log 2 +\log^2 2-9 \log 3+2 {\mathrm{Li}}_2\left(\frac{1}{4}\right)\right)\right]\,=\,I_{gq}^{(2)}\;,\nn\\[2mm]
I_{qg}^{(2)}&=&\frac{1}{2}\left[\frac{2}{3}+\varepsilon\left(\frac{67}{18}-12 \log 2+6 \log 3\right)\right]\;,\nn\\[2mm]
I_{gg}^{(2)}&=&2 C_A\left[-\frac{1}{\varepsilon}-\frac{11}{6} +\log 2+\frac{\varepsilon}{2} \left(-\frac{289}{18}+\pi^2
+30\log 2+\log^2 2-15\log 3+2{\mathrm{Li}}_2\left(\frac{1}{4}\right)\right)\right]\;,\nn\\
\eea
with the value ${\mathrm{Li}}_2(1/4)= 0.267653\ldots$ of the Dilogarithm function. Note that these integrals differ from
the corresponding ones for the cone or $k_t$ algorithms given in~\cite{jet1,jet2}, although the leading terms for
$\varepsilon\to 0$ are always the same. One can show that for arbitrary $n$ 
\bea\label{Inn}
I_{qq}^{(n)}&=&C_F\left[-\frac{1}{\varepsilon}-\frac{3}{2} +\log(n)+{\cal O}(\varepsilon)\right]\,=\,I_{gq}^{(n)}\;,\nn\\[2mm]
I_{qg}^{(n)}&=&\frac{1}{2}\left[\frac{2}{3}+{\cal O}(\varepsilon)\right]\;,\nn\\[2mm]
I_{gg}^{(n)}&=&2 C_A\left[-\frac{1}{\varepsilon}-\frac{11}{6} +\log(n)+{\cal O}(\varepsilon)\right]\;,
\eea
to which we will return later.

We now turn to the terms $d\sigma_{j(k)},d\sigma_{k(j)}$ in~(\ref{deco}). These are defined in such a way that they subtract 
the contribution where parton $j$ forms the jet, but parton $k$ is so close that it should normally be included in the jet 
(or vice versa). It is useful to introduce the pair mass $m$ of partons $j$ and $k$. Note that this is not the jet mass since 
for $d\sigma_{j(k)}$ only parton $j$ forms the jet. In terms of $m$ we find from Eq.~(13) of~\cite{jet1}  
\beq
\label{eq:scaps6}
\frac{d\sigma_{j(k)}}{dvdw}\,=\,\frac{\alpha_s}{\pi}\,{\cal N}_{ab\to K}(v,w,\varepsilon)\,
v\,(z_0(1-z_0))^{-\varepsilon}P_{jK}^{<}(z_0)
\,\int_0^{m^2_{{\mathrm{max}}}} \frac{dm^2}{m^2} \,m^{-2 \varepsilon},
\eeq
where $z_0=1-v+vw$ is the relevant variable for the splitting $K\to jk$. 
Again, implementation of the jet algorithm boils down to the determination
of the upper limit on the $m^2$ integration. The assumption that only parton $j$ forms 
the jet although really partons $j$ and $k$ should form it together, leads to the condition
\begin{align}
J^{(n)}(P_\J = p_j) < J^{(n)} (P_\J+ p_k)\label{condj(k)}\;.
\end{align}
This defines the subtraction $d\sigma_{j(k)}$. Using the NJA the relation becomes
\be
(P_\J^\perp)^n < (P_\J^\perp+p_k^\perp)^n\left(1-\beta\frac{m^2}{(P_\J^\perp+p_k^\perp)^2}\right)\;.
\ee
Again using the NJA, we have $p_k^\perp=P_\J^\perp (1-z_0)/z_0$ and hence obtain
\be
m^2 < \frac{(P_\J^\perp)^2}{\beta\,z_0^2}\left(1-z_0^n\right)\,=\, \frac{(P_\J^\perp)^2\left(1-(1-v+vw)^n\right)}{\beta\,(1-v+vw)^2}\;.
\ee
The right-hand-side is $m^2_{{\mathrm{max}}}$ to be used in Eq.~(\ref{eq:scaps6}). We therefore find
\beq
\label{eq:scaps6a}
\frac{d\sigma_{j(k)}}{dvdw}\,=\,\frac{\alpha_s}{\pi}\,{\cal N}_{ab\to K}(v,w,\varepsilon)\,v\,P_{jK}^{<}(z_0)\,
\left(-\frac{1}{\varepsilon}\right)\,\left( \frac{(P_\J^\perp)^2(1-z_0)\left(1-z_0^n\right)}{\beta z_0}\right)^{-\varepsilon}\;.
\eeq
The poles generated by this expression may be extracted in the standard way. If $P_{jK}^{<}$ is a non-diagonal 
splitting function one may simply expand in $\varepsilon$. If it is a diagonal one, it has a term $2C_j/(1-z_0)$ 
(where $C_q=C_F$ and $C_g=C_A$) that gives rise to a double pole. To see this, we observe that 
\beq\label{id1}
(1-z_0^n)^{-\varepsilon}\,=\,n^{-\varepsilon}(1-z_0)^{-\varepsilon}\left[1+{\cal O}((1-z_0)^2)\right]\;.
\eeq
We now use the identity
\beq\label{pid}
2C_j(1-z_0)^{-1-2\varepsilon}\,=\,C_j\left[-\frac{1}{\varepsilon}\delta(1-z_0)+\frac{2}{(1-z_0)_+}-4 \varepsilon
\left(\frac{\log(1-z_0)}{1-z_0}\right)_++{\cal O}(\varepsilon^2)\right]\;,
\eeq
where the ``plus''-distribution is defined in the usual way. Since $\delta(1-z_0)=\delta(1-w)/v$, the
pole term precisely matches those in Eqs.~(\ref{In1})--(\ref{Inn}). Keeping in mind that there is always
an overall factor $1/\varepsilon$ (see~(\ref{eq:scaps5a}),(\ref{eq:scaps6a})), this means that double poles
will cancel in the calculation. Furthermore, the factor $n^{-\varepsilon}$ in~(\ref{id1}) produces a
term $C_j\delta(1-z_0)\log(n)$ which matches the terms $\propto\log(n)$ in the finite parts of $I_{qq}^{(n)}$, 
$I_{gg}^{(n)}$ in~(\ref{Inn}). Thus, these terms will cancel as well in the difference $d\sigma_{jk}-d\sigma_{j(k)}$. 
Any remaining pole terms are independent of $n$ and are removed by the subtraction of final-state
collinear singularities in $d\sigma_{j(k)}$, as described in~\cite{jet1}. We have therefore shown that the
jet cross section defined according to~(\ref{Jdefn}) is infrared-safe at NLO for arbitrary $n$. 

The remaining task is to combine all pieces according to~(\ref{deco}). This proceeds as described 
in~\cite{jet1,jet2}. It is useful to compare the structure of the final result to that of the cross sections 
for the cone or $k_t$ algorithms in the NJA. For the latter, one has for a given partonic channel
\beq\label{B1}
d\sigma_{ab\to {\mathrm{jet}}X}\,=\,{\cal A}_{\,ab}\log (R^2) +{\cal B}_{ab}^{{\mathrm{\,algo}}}+{\cal O} (R^2)\;,
\eeq
where the ${\cal A}_{\,ab}$ are the same for both types of algorithms, but the ${\cal B}_{ab}^{{\mathrm{\,algo}}}$ 
depend on the algorithm, i.e. ${\cal B}_{ab}^{{\mathrm{\,cone}}}\neq {\cal B}_{ab}^{\,k_t}$. For the $J_{E_T}$ 
algorithms defined by maximizing the function $J^{(n)}$ in Eq.~(\ref{Jdefn}) we instead have 
from~(\ref{eq:scaps6a})--(\ref{pid})
\beq
d\sigma_{ab\to {\mathrm{jet}}X}\,=\,{\cal A}_{\,ab}\log \left( \frac{\left(1-z_0^n\right)}{\beta z_0(1-z_0)} \right) +
{\cal B}_{ab}^{(n)}+{\cal O} (1/\beta)\;,
\eeq
where again the ${\cal A}_{\,ab}$ are the same as for the other algorithms and the ${\cal B}_{ab}^{(n)}$ are all 
different and also differ from ${\cal B}_{ab}^{{\mathrm{\,cone}}}$ and ${\cal B}_{ab}^{\,k_t}$. For the most 
important case $n=1$ we have, using $z_0=1-v+vw$, 
\beq\label{B2}
d\sigma_{ab\to {\mathrm{jet}}X}\,=\,{\cal A}_{\,ab}\log \left( \frac{1}{\beta (1-v+vw)} \right) +
{\cal B}_{ab}^{(1)}+{\cal O} (1/\beta)\;.
\eeq
In any case, one can see that there is a simple correspondence between logarithms of $R^2$ for
the cone and $k_t$ algorithms and logarithms of $1/\beta$ for the $J_{E_T}$ ones.
The implementation into the numerical code of~\cite{jet1,jet2} is thus relatively straightforward.

\section{Phenomenological results \label{Pheno}}

We now present a few phenomenological results for the NLO jet cross section for the new class of algorithms. 
For our studies we consider $p\bar{p}$ collisions at center-of-mass
energy $\sqrt{S}=1.96$~TeV at the Tevatron, and $pp$ collisions at $\sqrt{S}=7$~TeV at the LHC.
For Tevatron we choose a jet rapidity interval $|\eta_\J|\leq 0.4$, while for the LHC we use three different 
bins in rapidity, $|\eta_\J|\leq 0.5$, $2\leq |\eta_\J|\leq 2.5$, and $4\leq |\eta_\J|\leq 4.5$.
We use the CTEQ6.6M parton distributions~\cite{cteq66} and the renormalization/factorization scale
$\mu=P_\J^\perp$ throughout. All our calculations presented here are carried out in the context of the 
NJA. We note that in Ref.~\cite{jet2} detailed comparisons of the NLO jet cross
sections obtained within the NJA and obtained with a full NLO Monte-Carlo integration code, respectively, were
performed, both for the cone and for the $k_t$ algorithm. These comparisons showed that the NJA is
very accurate for values $R=0.4$ of the jet parameter, for the kinematics of interest in our present
study. Moreover, {\it ratios} of cross sections obtained in the NJA typically match full NLO ones
even better. We therefore always present our results for the new algorithms relative to the one
for the $k_t$ algorithm in the NJA with $R=0.4$. 
\begin{figure}[t]
\centering
\vspace*{-0.cm}
\hspace*{-0.2cm}
\epsfig{figure=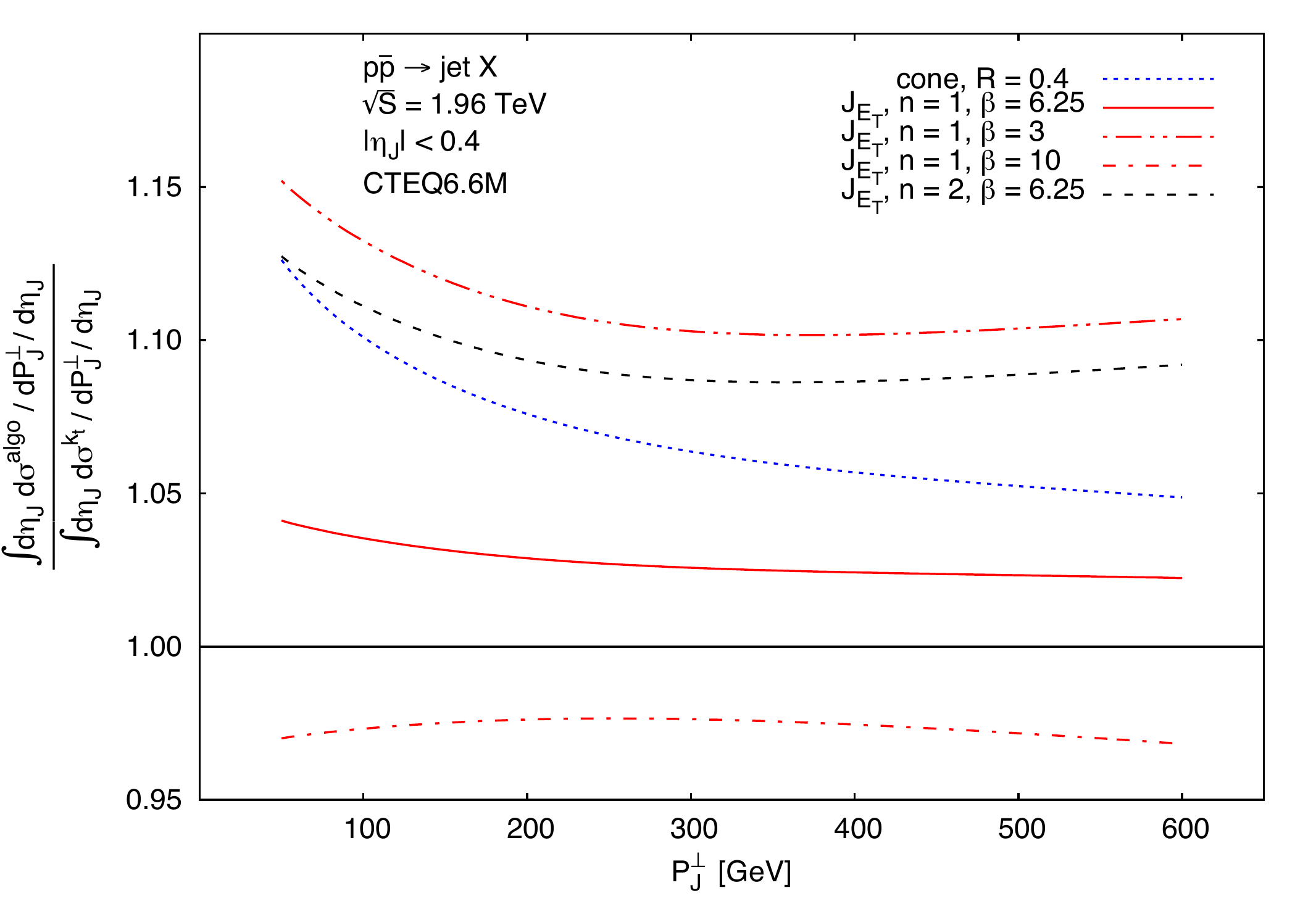,width=0.9\textwidth}

\vspace*{-0cm}
\caption{\sf Comparison of jet cross sections in the NJA for Tevatron kinematics. All results are shown relative
to the one for the $k_t$ algorithm with $R=0.4$. \label{fig:Tevatron}}
\vspace*{0.cm}
\end{figure}

Figure~\ref{fig:Tevatron} shows our results for Tevatron kinematics. The solid line is the result
for the new algorithm with $n=1$, using $\beta=6.25$ which equals the value $1/R^2$ 
used for the baseline calculation of the cross section for the $k_t$ algorithm. One can see that the
result is relatively close to the one for the $k_t$ algorithm in this case, indicating that the 
difference induced by the non-logarithmic pieces ${\cal B}_{ab}^{\,\mathrm{algo}}$ in~(\ref{B1}),(\ref{B2}) 
is relatively small. To explore the dependence on $\beta$ we also present results for $\beta=3$ and
$\beta=10$ (dash-dotted lines), which are higher or lower, respectively, than the one for the $k_t$ algorithm. 
Empirically, one finds that a value $\beta\approx 1.25/R^2$ leads to a ratio very close to unity. Such a
finding is expected when the cross section has the form given in Eqs.~(\ref{B1}),(\ref{B2}). (We recall in this
context that Ref.~\cite{catani} observed that the cone and $k_t$ algorithms lead to similar results when
$R_{k_t}\approx 1.35 \,R_{\mathrm{cone}}$). The dashed line in Fig.~\ref{fig:Tevatron} shows the
result for the $J_{E_T}$ algorithm with $n=2$ and $\beta=6.25$. It is significantly higher 
(by about 10\%) than the baseline one for the $k_t$ algorithm. In fact, it is closer to that for the 
cone algorithm with $R=0.4$, which is also shown in the figure by the dotted line. 

\begin{figure}[t]
\centering
\vspace*{-0.cm}
\hspace*{-0.2cm}
\epsfig{figure=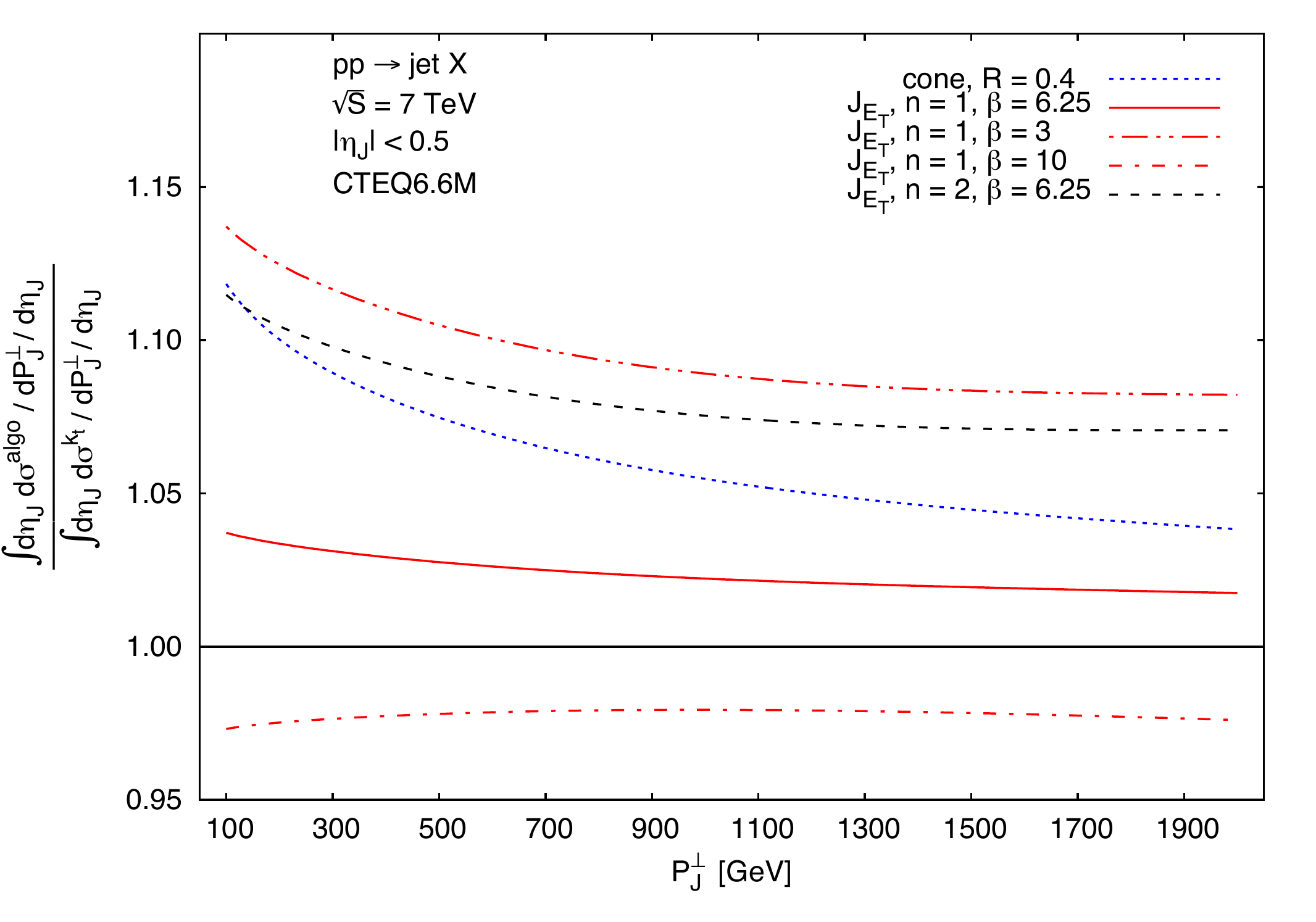,width=0.9\textwidth}

\vspace*{-0cm}
\caption{\sf Same as Fig.~\ref{fig:Tevatron} but for $pp$ collisions at $\sqrt{S}=7$~TeV. \label{fig:LHC}}
\vspace*{0.cm}
\end{figure}

\begin{figure}[t]
\centering
\vspace*{-0.cm}
\hspace*{-0.2cm}
\epsfig{figure=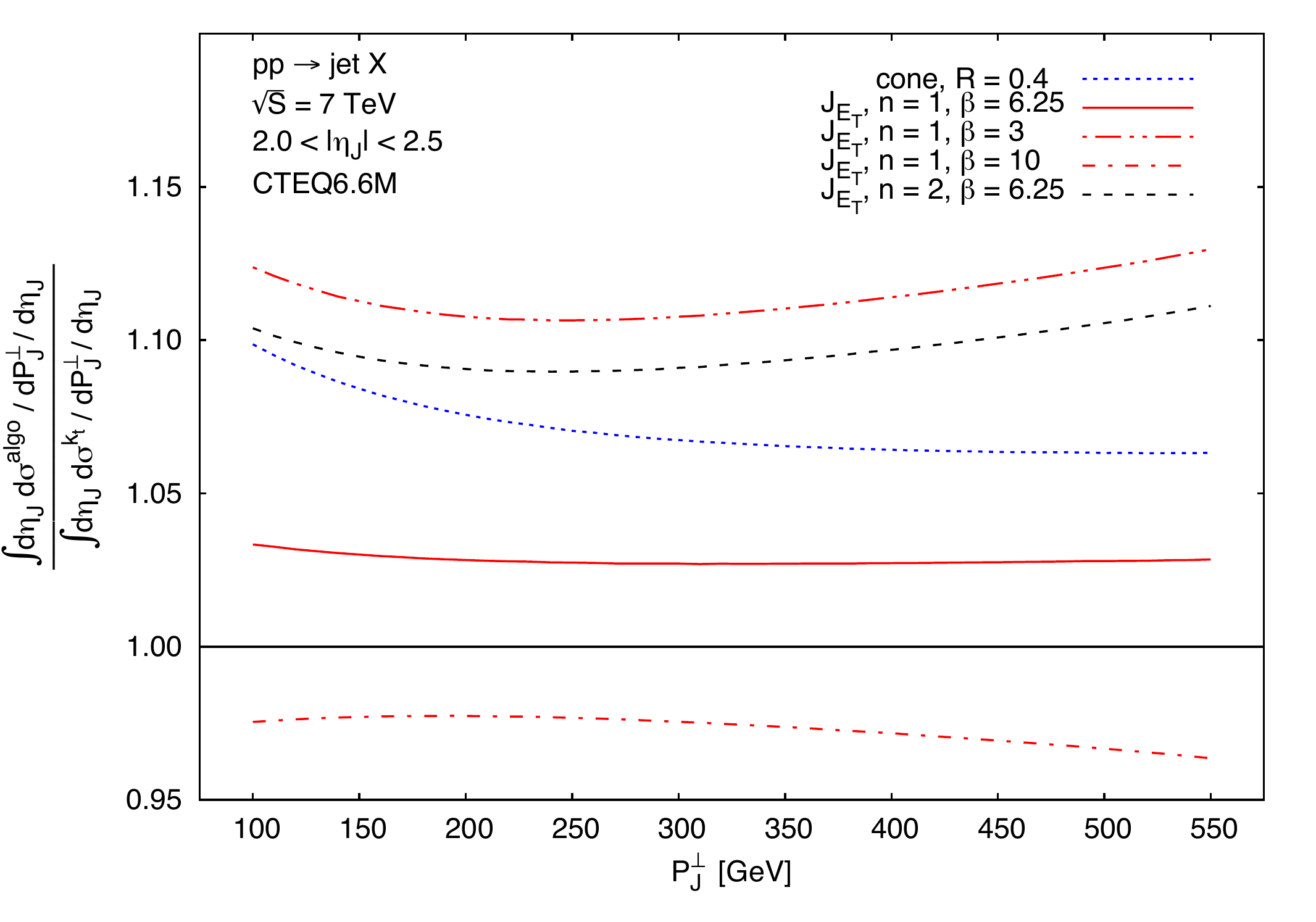,width=0.9\textwidth}

\vspace*{-0cm}
\caption{\sf Same as Fig.~\ref{fig:LHC} but for $2\leq |\eta_\J|\leq 2.5$. \label{fig:LHC2}}
\vspace*{0.cm}
\end{figure}

\begin{figure}[t]
\centering
\vspace*{-0.cm}
\hspace*{-0.2cm}
\epsfig{figure=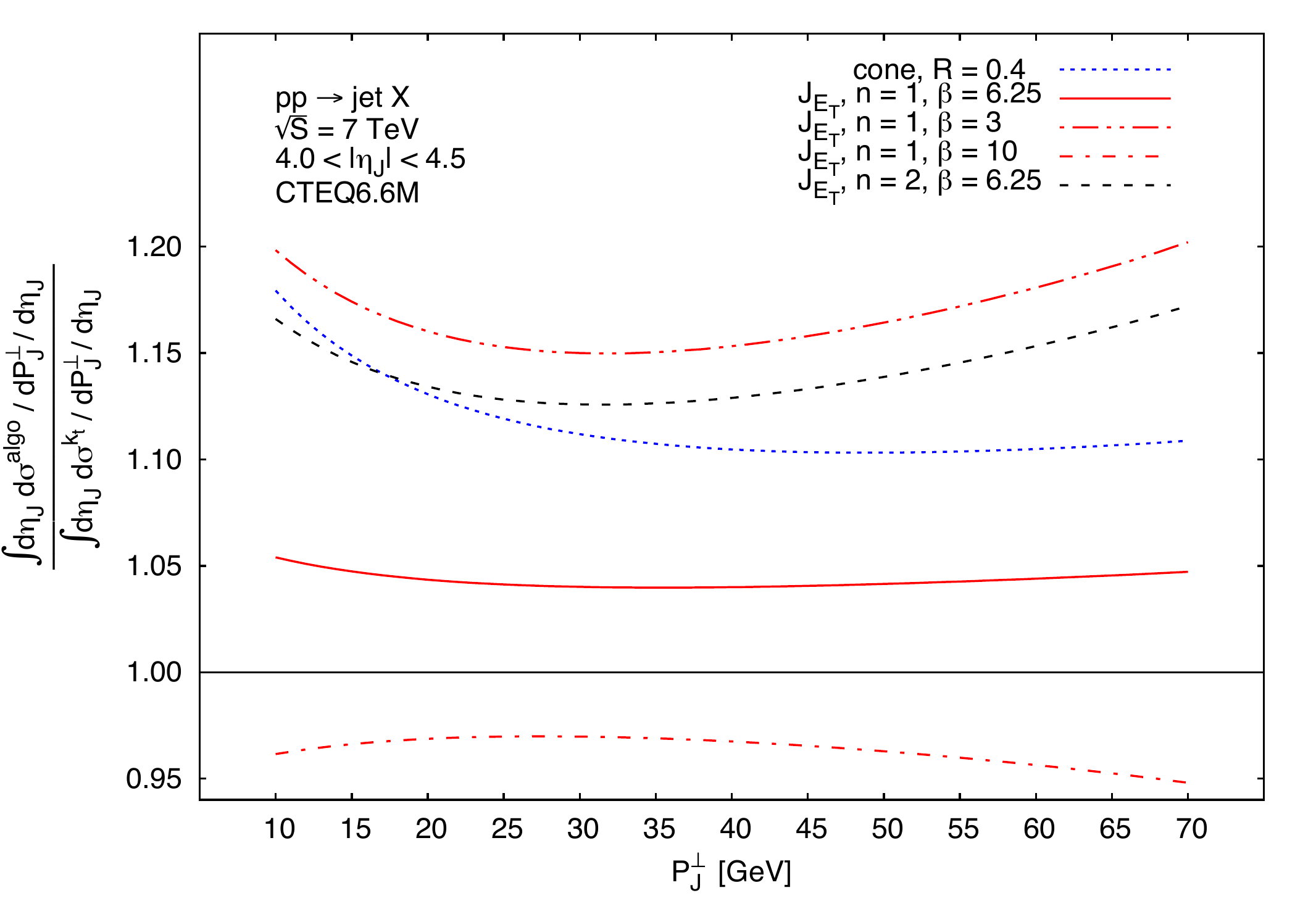,width=0.9\textwidth}

\vspace*{-0cm}
\caption{\sf Same as Fig.~\ref{fig:LHC} but for $4\leq |\eta_\J|\leq 4.5$. \label{fig:LHC4}}
\vspace*{0.cm}
\end{figure}

In Fig.~\ref{fig:LHC} we show our results for mid-rapidity jet production in $pp$ collisions at the LHC. 
As one can see, all features found for Tevatron conditions carry over to this case as well. This remains
essentially true also for jets produced at larger $|\eta_\J|$, as shown by Figs.~\ref{fig:LHC2} and~\ref{fig:LHC4}
for the cases $2\leq |\eta_\J|\leq 2.5$ and $4\leq |\eta_\J|\leq 4.5$, respectively.

\section{Conclusions \label{con}} 

We have presented a next-to-leading order calculation for single-inclusive jet production in hadronic
collisions, using the recently introduced $J_{E_T}$ algorithms to define jets. Our calculations have
been performed analytically, assuming that the produced jets are rather collimated. We have found that 
all singular contributions arising at intermediate stages of the NLO calculation cancel in the final answer,
which is a prerequisite for an infrared safe algorithm. Our numerical studies show that jets defined according
to the $J_{E_T}$ algorithms have a cross section that is overall rather close to those for the more
standard cone or $k_t$ algorithms. While future work will need to decide whether the $J_{E_T}$ 
algorithms offer any advantages over the standard ones in actual experimental jet studies, we think 
that our results are useful in assessing their theoretical status. Our analytical results will also be 
valuable for QCD resummation studies of jet production for the $J_{E_T}$ algorithms, 
cf.~\cite{deFlorian:2013qia}.

\section*{Acknowledgments} 

AM thanks the Alexander von Humboldt Foundation, Germany, for support through a Fellowship for Experienced Researchers.
This work was supported in part by the Institutional Strategy of the University of T\"{u}bingen (DFG, ZUK 63).




\newpage
\begin{thebibliography}{99}

\bibitem{soyez} {\it See:} G.~Soyez,
  Nucl.\ Phys.\ Proc.\ Suppl.\  {\bf 191}, 131 (2009)  [arXiv:0812.2362 [hep-ph]].
  
\bibitem{soyez1}
G.~P.~Salam and G.~Soyez,
  JHEP {\bf 0705}, 086 (2007)
  [arXiv:0704.0292 [hep-ph]].
  
\bibitem{ES}  S.~D.~Ellis and D.~E.~Soper,
  Phys.\ Rev.\ D {\bf 48}, 3160 (1993)
  [hep-ph/9305266].
  
\bibitem{catani}  
S.~Catani, Y.~L.~Dokshitzer, M.~H.~Seymour and B.~R.~Webber,
  Nucl.\ Phys.\ B {\bf 406}, 187 (1993).

\bibitem{css} M.~Cacciari, G.~P.~Salam and G.~Soyez,
  JHEP {\bf 0804}, 063 (2008)  [arXiv:0802.1189 [hep-ph]].

\bibitem{georgi} 
  H.~Georgi,
  arXiv:1408.1161 [hep-ph].
  
\bibitem{Bai:2014qca} 
  Y.~Bai, Z.~Han and R.~Lu,
  arXiv:1411.3705 [hep-ph].
  
\bibitem{Ge:2014ova} 
  S.~F.~Ge,
  arXiv:1408.3823 [hep-ph].
  
\bibitem{Stewart:2010tn} 
  I.~W.~Stewart, F.~J.~Tackmann and W.~J.~Waalewijn,
  Phys.\ Rev.\ Lett.\  {\bf 105}, 092002 (2010)
  [arXiv:1004.2489 [hep-ph]].
  
\bibitem{Aversa:1990ww} 
  F.~Aversa, L.~Gonzales, M.~Greco, P.~Chiappetta and J.~P.~Guillet,
  Z.\ Phys.\ C {\bf 49}, 459 (1991);  J.~P.~Guillet,
  Z.\ Phys.\ C {\bf 51}, 587 (1991).  
  
\bibitem{jet1} B.~J\"{a}ger, M.~Stratmann and W.~Vogelsang,
  Phys.\ Rev.\ D {\bf 70}, 034010 (2004)
  [hep-ph/0404057].

\bibitem{jet2} A.~Mukherjee and W.~Vogelsang,
  Phys.\ Rev.\ D {\bf 86}, 094009 (2012)
  [arXiv:1209.1785 [hep-ph]].
  
\bibitem{oldsca2}   F.~Aversa, P.~Chiappetta, M.~Greco and J.~P.~Guillet,
  Nucl.\ Phys.\ B {\bf 327}, 105 (1989).

\bibitem{jssv}
B.~J\"{a}ger, A.~Sch\"{a}fer, M.~Stratmann and W.~Vogelsang,
Phys.\ Rev.\ D {\bf 67}, 054005 (2003)  [hep-ph/0211007].

\bibitem{cteq66}
P.~M.~Nadolsky {\it et al.},
  Phys.\ Rev.\ D {\bf 78}, 013004 (2008)
  [arXiv:0802.0007 [hep-ph]].

\bibitem{deFlorian:2013qia} 
  D.~de Florian, P.~Hinderer, A.~Mukherjee, F.~Ringer and W.~Vogelsang,
  Phys.\ Rev.\ Lett.\  {\bf 112}, 082001 (2014)
  [arXiv:1310.7192 [hep-ph]].
    
\end{thebibliography}
\end{document}